\input epsf.tex
\input epsf.sty
\documentstyle[12pt]{article}
\begin{document}
\begin{titlepage}
\begin{center}
July 1995            \hfill CERN-TH 95-280 \\
        \hfill LBL-37559 \\
          \hfill    UCB-PTH-95/27 \\
         \hfill   hep-th/9511101 \\
 
\vskip .25in
 
{\large \bf Variational solution of the Gross-Neveu model:\\
Finite $N$ and renormalization}\footnote{This work was
supported in part by the Director, Office of
Energy Research, Office of High Energy and Nuclear Physics, Division of
High Energy Physics of the U.S. Department of Energy under Contract
DE-AC03-76SF00098 and in part by the National Science Foundation under
grant PHY-90-21139.}
 
\vskip .25in
C. Arvanitis$^\dagger$\footnote{c.arvanitis@ic.ac.uk. Supported by the EC
under H.C.M Grant No ERBCHBICT941235},
F. Geniet$^\ddagger$\footnote{geniet@lpm.univ-montp2.fr},
M. Iacomi$^\ddagger$\footnote{iacomi@lpm.univ-montp2.fr}, 
J.-L. Kneur$^{\star}$\footnote{kneur@mail.cern.ch. On leave
from Laboratoire de Physique
Math\'ematique, Universit\'e
Montpellier II-CNRS,
34095 Montpellier c\'edex 05.}
and
A. Neveu$^{\ddagger,\S}$\footnote{neveu@lpm.univ-montp2.fr. 
 On sabbatical leave
after Sept. 1$^{\hbox{st}}$ 1994 from Laboratoire de Physique Math\'ematique,
Universit\'e de Montpellier II-CNRS, 34095 Montpellier c\'edex 05}
\\
{$^\dagger$ \em Physics Department,
Imperial College,  \\
Theoretical Physics Group, \\
Prince Consort Rd.,
 London SW7 2BZ, UK} \\ 
\vskip .3 cm
{$^\ddagger$ \em Laboratoire de Physique Math\'ematique\footnote{Laboratoire
associ\'e au Centre National de la Recherche Scientifique.}
\\
Universit\'e Montpellier II-CNRS
F-34095 Montpellier c\'edex 05.}\\
\vskip .3 cm
{$^{\star}$ CERN,
\em Theoretical Physics Division \\
CH-1211 Geneva 23 Switzerland} \\
\vskip .3 cm
{$^{\S}$ \em 
Theoretical Physics Group\\
    Lawrence Berkeley Laboratory\\
      University of California,
    Berkeley, California 94720}
\end{center}
\vskip .5in
\newpage 
\begin{abstract}
We show how to perform systematically improvable
variational calculations in the $O(2N)$ Gross-Neveu model for
generic $N$, in such a way that all infinities usually plaguing such
calculations
are accounted for in a way compatible with the perturbative
renormalization group .
The final point is a general framework for the calculation of 
non-perturbative
quantities like condensates, masses etc$\ldots$, in an asymptotically free
field theory.
For the Gross-Neveu model, the numerical results obtained from a ``2-loop''
variational calculation are in a very good agreement with exact quantities
down to low values of $N$.
\end{abstract}
\end{titlepage}
%
%
%
%

 
 
\newpage
\renewcommand{\thepage}{\arabic{page}}
\setcounter{page}{1}
 
\def\ss{\vskip10pt}
\def\hh{\hskip10pt}
\def\beq{\begin{equation}}
\def\eeq{\end{equation}}
\def\beqa{\begin{eqnarray}}
\def\eeqa{\end{eqnarray}}
\def\ds{\displaystyle}
\newcommand{\ggt}{\tilde\Gamma}
\newcommand{\mub}{\mu e^{\frac{1}{2}(-\gamma_{E}+\ln 4\pi)}  }
\newcommand{\RG}{renormalization group}
\newcommand{\eps}{\varepsilon}
\newcommand{\mo}{m_{0}}
\newcommand{\go}{g_{0}}
\newcommand{\Lam}{\Lambda_{\overline{MS}}}
\renewcommand{\theequation}{\thesection.\arabic{equation}}
\setcounter{equation}{0}
\setcounter{figure}{0}
\setcounter{table}{0}
\setcounter{footnote}{0}
\parskip 10pt
\vfill\eject
 
\section{Introduction.} \label{intro}
 
It has been advocated for a long time (\cite{CAS}-\cite{SOLO})
that the convergence of conventional perturbation
theory may be improved by a variational procedure in which
the separation of the
action into ``free" and ``interaction" parts
 is made to depend on some set of auxiliary parameters.
The results obtained  by expanding to finite order
 in this ``redefined" perturbation
series are optimal  in regions  of the  space of auxiliary
parameters where they are least sensitive to these parameters.
This is reasonable: such regions are those expected to
resemble the most to the complete answer, where there should
simply be {\em no} dependence on the auxiliary parameters.
This intuitive argument can be made more precise, and
there has recently appeared
strong evidence that this optimized perturbation
theory may indeed lead to a rigorously convergent
series of approximations even in strong coupling cases.
In particular, the convergence of this variational-like procedure has been
rigorously established in the case of zero
and one dimensional ``field theories"~\cite{JONE}.
 
Having these rigorous proofs, it is very tempting
to try to apply these methods in renormalizable quantum field theory.
The main obstacle which one should overcome in this case is the compatibility
of the
variational expansion with the usual renormalization program of the theory.
Indeed there are very few  field-theoretical cases where the method
has been tested and gave non-trivial
results. Most of these cases concern the $\phi^4$
theory, which is unfortunately free in four dimensions, and the
effective potential of the large-$N$ Gross-Neveu
(GN) model~\cite{OKO,MOS,PIN,SOLO}.
 
In this paper, in the finite-$N$ GN model,
we shall show how one can build variational expansions which
are compatible with perturbative renormalization and give
non-trivial results even for low values of $N$.
Our approach can be in principle extended  to more general theories,
like QCD, as a new non-perturbative approach to the calculation
of (dynamical) chiral symmetry breaking parameters~\cite{AGKN}.

We shall first take a closer look at the large-$N$~\cite{AGN}
 limit and from there 
infer the procedure for the case of arbitrary $N$.
Taking the fermion mass
as a variational
parameter we can obtain exact 
results in the large-$N$ limit of the $O(2N)$
GN 
model, but as we shall see,  
in the finite-$N$ case it seems necessary to go to infinitely
high
perturbative order just to obtain the right renormalization 
group (RG) behavior.
We will see how to overcome this problem and interpret it 
in order to automatically include the correct renormalization group
properties, obtaining finite answers as the space-time dimension goes to 2.
This turns out to be a relatively simple exercise in renormalization theory,
taken from an unusual point of view.
At the end, we obtain a variational estimate of 
the vacuum energy and the mass
gap, as a function of $\Lambda_{\overline{MS}}$, whose accuracy increases
with the order of the perturbative expansion.
Exact results for these two quantities are known~\cite{FNW,ZAM}
and provide a test for the accuracy of the method.
Note that our method also has definite connections
with
previous derivations
of a mass gap in the framework of the (non-linear) sigma model
\cite{BREZINN,JOLINI}. In particular our way of using the
RG properties
to relate the mass gap to the basic scale,
$\Lam$, is quite analogous with the one in the latter reference,
although the subsequent method to extrapolate
the pure RG results to the actual mass gap values is completely different.
Moreover, as above mentioned the essential novelty in the
present approach is to clarify the link between RG results and the
variational expansion, allowing a treatment of the latter
that is consistent with infinities and therefore much more adaptable
to other theories.

In section~\ref{bare}, we explain the basic
idea and ingredients of our method, 
using only the first coefficients of
the \RG\   $\beta(g)$ and $\gamma(g)$ functions in order to keep
the discussion clear. In section \ref{highord}, 
we show how to generalize, formally, to 
arbitrary higher orders the previous construction, 
and derive explicit formulas for the fermion mass and the vacuum energy
density using the presently known maximal perturbative order, i.e.
the 
three-loop \RG\  functions. 
In section~\ref{alternative}, we construct an alternative Ansatz, 
starting
directly from renormalized quantities, which is shown to differ 
in general from the previous one by a specific renormalization scheme
change, and lead to essentially
similar results. It also gives a more transparent construction, 
in particular more appropriate to generalization to higher
orders and other theories. 
Finally some technicalities are treated in two Appendices. 
 
\section{One-loop renormalization group properties} \label{bare}

Let us start from the $O(2N)$
GN Lagrangian~\cite{GN}, modified with a non-zero fermion mass as follows
\beq
L_{GN}(\mo ,\go, x) = \bar \Psi (i \gamma_\mu \partial^\mu -\mo ) \Psi
+x\; ( \frac{\go^2}{2} (\bar \Psi \Psi)^2 +\mo \bar \Psi \Psi ) ;
\label{LGN}
\eeq 

\noindent where we introduced for convenience a parameter $x$,
interpolating between the {\it free}, massive Lagrangian for $x = 0$, and
the usual massless GN interaction Lagrangian, for $x=1$ (summation
over the $2N$ component of fermions $\Psi$ is implicitely 
understood~\footnote{We consider the $O(2N)$ case only for
convenience of comparison with the results of ref.\cite{FNW}, but in fact
there are a priori no limitations in our method to consider $N$ 
half-integer as well, i.e. corresponding to the usual $O(N)$ model.}). 
As is well known, one can solve the model exactly in the large N limit 
and obtain a non-trivial mass gap in the massless limit, i.e. for 
$x \to 1$ in the present context. Now how can we treat 
the theory as defined by (\ref{LGN}) for arbitrary $N$? A direct application
of the arguments developed in the introduction 
would consist of doing perturbation theory with respect to the
reorganized Lagrangian above, 
i.e. expanding the above terms to some order in $x$. 
One would then look for extrema with respect to $m$, supposedly approximating
the unknown exact result for $x\to 1 $, which by definition does not depend
on $m$. Lessons from the anharmonic oscillator case leads one to 
expect a systematic 
improvement when going to higher orders in the expansion in $x$, the better
approximation being assumed for the flattest extrema structure.

Unfortunately, one immediately encounters a number of obstacles. 
Most importantly, before accessing to any physical quantity
of interest for such an optimization the theory has to be
renormalized, and there is an unavoidable mismatch
between the expansion in $x$ as
introduced above and the ordinary perturbative expansion as dictated by the
necessary counterterms, as we shall see explicitely in section 2.2. 
Independently of that, 
even in the most optimistic case where
arbitrary orders of perturbation theory would be known 
(which is only true in the large $N$ case),
it is easily seen that at any {\em finite} order 
one only recovers a trivial result 
for $x\to 1$ (i.e. $ m_0 \to 0$), which
is the limit we are however interested in to identify a non trivial
mass gap. 
 
Actually these facts can be circumvented by advocating  a certain
resummation Ansatz, whose main ingredient exploits the renormalization group
invariance of the theory and analytic continuation properties 
of the (arbitrary) mass parameter $m$~\footnote{More precisely, of an
arbitrary parameter related to the original Lagrangian mass $\mo$
in a way to be specified next.}, 
which at the end defines a mass gap as an 
integral transform as we shall derive below. 
This (renormalization group invariant) 
Ansatz is exact in the
large $N$ limit, where it can be shown to resum the $x$ dependence exactly. 
In the finite
N case 
it is only exact as far as the leading 
(or next-to-leading)
renormalization group behavior is concerned. But the important point
is that it 
also provides a non-trivial resummation of the perturbative serie in $x$, 
in the sense that it gives a finite and non-zero result in the $m \to 0$
limit. This is expected to be a sufficiently good ``trial" Ansatz 
for a subsequent 
optimization with respect to $m$, as will be motivated
below. 

\subsection{The mass gap in the large $N$ limit}
To be more concrete let us 
start with the expression of the 
one-loop mass and coupling counterterms of the model, 
expressed with the renormalized parameters $m$ and $g$ related to
$\mo$ and $\go$ by :
\beqa
\mo = m Z_{m}= \frac{m}{[1-\frac{g^{2}(N-1)}
{\pi\eps}]^{\frac{N-1/2}{N-1}}}
\hh , \nonumber \\
\go^{2} = g^{2}\mu^{-\eps}Z_{g}=
\frac{g^{2}\mu^{-\eps}}
{1-\frac{g^{2}(N-1)}{\pi\eps}}
\hh 
\label{zmzg} 
\eeqa
where $\mu$ is the arbitrary scale introduced by 
dimensional regularization.
Note that those one-loop expressions are exact in the $N \to \infty$ limit,
where it can be shown that there are
no other leading $N$ corrections. For finite $N$ this one-loop RG 
dependence generates however only the leading terms in $1/\epsilon$,
to all orders.  

Consider now the following expression for a (bare) resummed mass,
as inspired by the latter RG properties:
\beq
m_F(\mo) = \frac{\mo}{(1- 2(N-1) \go^2 
\ggt \mu^{-\eps} m^{\eps}_F(\mo) )^{\frac{N-1/2}{N-1} } }
\label{mf1}
\eeq
(where $\ggt \equiv \Gamma[-\eps/2]/(4\pi)^{1+\eps /2}$). 
Eq.~(\ref{mf1}) is obviously renormalization group invariant since expressed  
only in terms of 
the bare parameters $\mo$ and $\go$, and can be expressed
in terms of the renormalized parameters $m(\mu)$ and $g(\mu)$ 
with the help of
eq.~(\ref{zmzg}) above.  
When this is done, one obtains for $m_{F}$ an expression which can be
recursively expanded in powers of $g^2(\mu)$ and, as easily checked, 
is finite at each order as
$\eps\rightarrow 0$, at fixed $g^2(\mu)$ and $m(\mu)$, 
thanks to the recursive
dependence in $m_F$ in eq. (\ref{mf1}).
Explicitely one obtains 
\beq
m_F(\bar m) = \bar m \;[ 1 +\frac{(N-1)}{\pi} \bar g^2 \; 
\ln [\frac{m_F(\bar m)}
{\bar\mu}]\; 
]^{-\frac{N-1/2}{N-1}}\; ,
\label{mfren1}
\eeq
where the $\overline{MS}$ scheme scale, 
${\overline \mu} = \mub$ was introduced
for convenience, and 
$\bar g \equiv g(\bar\mu)$, $\bar m \equiv m(\bar\mu)$. 
Eq.~(\ref{mfren1}) resums the leading logarithmic
dependence in $\bar m$, and in the $N \to \infty$ limit there are no other
corrections. Using the recursivity one
can rewrite (\ref{mfren1}) 
identically as 
\beq
m_F (m'' ) = \Lam \; \frac{m''}{f(m'')^{\frac{N-1/2}{N-1}}}\; ;
\eeq
where the scale-invariant, dimensionless mass parameter: \\
$m'' \equiv (\bar m/\Lam)\; [(N-1) \bar g^2/\pi]^{-\frac{N-1/2}{N-1}}$ 
has been
introduced for convenience
and is related to $f$ as 
\beq 
f(m'') +\frac{N-1/2}{N-1}\;\ln f(m'') \equiv \ln m''
\label{Frec1}
\eeq
while 
$\Lam \equiv \overline \mu ~exp[-\pi/((N-1)g^2(\overline \mu ))\;]$ 
is the $\overline{MS}$ 
RG-invariant basic scale. \\ 
Now from (\ref{Frec1}), it is easily seen
that for
$m'' \to 0$ (i.e. $m \to 0$), $f(m'') \simeq (m'')^{\frac{N-1}{N-1/2}}$, 
so that one recovers the well
known result,
\beq 
 m_F = \Lam 
\nonumber
\eeq 
for the mass-gap in the large $N$ limit.
\subsection{Usual plague of the $x$ expansion and its cure}

Up to now we have not made any use of the expansion in the 
``variational" parameter $x$ and indeed, as expected, that was not 
necessary to obtain the correct 
result for the mass gap in the $N \to \infty$ limit. 
Since the large $N$ result is a good consistency check, and
the perturbation in $x$ will play a definite r\^ole later on in the 
more complicated arbitrary $N$ case, let us see what is happening
when considering the $x$ expansion. 

Introducing the $x$ expansion parameter in (\ref{LGN}) 
formally amounts to substitute
$\mo \to \mo (1-x)$, $\go^2 \to \go^2 x $ in any (bare)
expression, in particular in (\ref{mf1}). One however 
soon realizes
that (\ref{mf1}) is then no longer finite for $ \epsilon \to 0$
at any order, apart for $x =1$, in which case only the trivial result
$m_F = 0$ is recovered,
as announced. 
Indeed the resummed expression (\ref{mfren1}) resulted from the usual 
RG properties and has no reason to be compatible with the way the
perturbative $x$ expansion is introduced. 
The cure to that problem is to resum 
 the series
$m_F(x) \equiv \sum^\infty _{x=0} a_q x^q $, in a different way:
this is made possible by analytic continuation in $x$, 
and an adequate contour integral can be shown~\cite{AGN} to resum exactly 
(in the large $N$ limit) the
above series, as  
explained in some details in Appendix A. Apart from technical details,
the main effect of the resummation is to cancel the factor of $(1-x)$
which arises from the substitution applied to $m_0$ and was responsible
for the trivial result at any arbitrary order for $x \to 1$.
The net result gives the mass gap as a 
specific contour integral expression:
\beq
\left.
\begin{array}{l}
m_{F}(\mo') = \ds{\frac{\mo'}{2i\pi}\oint\frac{du{\rm e}^{u}}{f_{1}^
{\frac{N-1/2}{N-1}}(u)} } \\
\mbox{with} \\
f_{1}(u)  =  1 - 2(N-1)\go^2 \ggt \; (\mo'u)^{\eps} \; {[f_{1}(u)]}
^{-\frac{N-1/2}{N-1} \eps}
\label{mff}
\end{array}
\right\}
\eeq
where
$u \equiv q(1-x)$ has been introduced as an appropriate change of
variable to analyse the $x \to 1$, $q \to \infty$ limit of the
$q^{th}$-order expansion of $m_F(x)$,
$\mo'$ is obtained by the rescaling $\mo \equiv q\;\mo'$,
and the integration contour runs counterclockwise along the cut negative real
axis.
 
The other nice thing about equation~(\ref{mff}) is, 
that it has a smooth limit
as $\eps\rightarrow 0$, at fixed renormalized $ m'$ and $g^2$, where 
$m'$ is obviously related to $m'_0$ as in (\ref{zmzg}).
Indeed in the limit $\eps\rightarrow 0$, we have :
\beq
\left.
\begin{array}{l}
m_{F}(m')=\ds{\frac{m'}{2i\pi}\oint\frac{du{\rm e}^{u}}{[f'_{1}(u)]^
{\frac{N-1/2}{N-1}}}}  \\
\mbox{with:}  \\
f'_{1}(u)=1+\ds {\frac{g^{2}(N-1)}{\pi} \; ( \; \ln  \frac{m'u}{\overline \mu}
-\frac{N-1/2}{N-1} \ln f'_{1}\;)}\;.
\end{array}
\right\}
\label{MF}
\eeq
As expected, this can be expressed in terms of the dimensionless
parameter $ m''$, given by
\beq
m''= \ds{\frac{m'}{\Lam \; 
[\frac{(N-1)g^2}{\pi}]^{\frac{N-1/2}{N-1}}}} \hh ,
\label{msecond}
\eeq

which is invariant under the renormalization group of the
{\em massive} theory i.e.
\beq
\Bigl\{\mu\frac{\partial}{\partial\mu} \; - \; \frac{g^{2}(N-1)}{\pi}g
\frac{\partial}{\partial g} \; - \;
\frac{g^{2}(N-1/2)}{\pi}m'\frac{\partial}{\partial m'}\Bigr\} \; m''=0 \hh .
\eeq
 
After these rescalings one obtains:
\beq
m_{F}(m'')=\ds{ \Lam \;
 \frac{m''}{2i\pi}\oint\frac{du{\rm e}^{u}}{f^
{\frac{N-1/2}{N-1}}(u)}} \hh ,
\label{MFF}
\eeq
with\footnote{Note that $f(u)$ is related to $f'_1(u)$ in (\ref{MF})
as $f(u) \equiv \frac{\pi}{(N-1)g^2} f'_1(u)$. }
\beq
f(u)=\ln(m''u)-\frac{N-1/2}{N-1}\ln f(u) \hh .
\label{furec}
\eeq
 
This conceptually very simple expression defines a function $m_{F}(m'')$,
which should be studied, looking for extrema, and/or its value
at $ m''=0 $, as explained in references~\cite{AGN,BGN1}. In fact,
as far as the large $N$ limit is concerned 
the optimization is trivial, as one may expect: from (\ref{furec})
one can find the series expansion of $f(u)$ near the origin as
$f(u) \simeq (m''u)^{\frac{N-1}{N-1/2}}$, from which it is
immediate that the expression
(\ref{MFF}) has a simple pole at $u =0$,
whose residu simply gives $m_F = \Lam$. Now however,
 the main interest of
the above construction is 
that the resummation of the $x$ expansion remains
valid, when considering the generalization at higher
RG orders of the different expressions, 
and when one includes as well the non-RG perturbative corrections
which are present at the next order, as we shall see.  
Remark at this point that all the complexities of 
the renormalization procedure
are hidden in the definition of the function $f(u)$ in (\ref{furec}).
We note that once the pure number $m''$ is fixed, for example at an extremum
of $m_{F}(m'')$, the corresponding value of $ m_{F}$ is an invariant
of the renormalization group of the {\em massless}
theory, as is evident from equation~(\ref{MFF}).
\section{Higher order results}\label{highord} 
\setcounter{equation}{0}
\subsection{Generalized Ansatz}
For finite $N$, and at a given order in $g^2$, one knows $Z_{m}$ and
$Z_{g}$ from the knowledge of the \RG\ functions
$\beta(g)$ and $\gamma(g)$ at the appropriate order in $g^2$.
Hence we can ask the question :
 
Can we guess for $m_{F}$ (or any other physical quantity) an expansion
in powers of $\go^{2} \mo^{\eps}$ which is finite when $\eps\rightarrow 0$
at fixed $g^2$ and $m$ and has a power expansion in $g^2$ ?
 
The exact expression for $m_{F}$ is of course an answer, but certainly
not the only one, to this simple purely mathematical question.
What we are after is a {\em class} of answers, and a general
principle which  will enable us to pick in this class the appropriate answer
which reproduces the results of the calculation up to a given order
in the loop expansion, {\em i.e.} in the expansion of $m_{F}$ in powers
of $\go^{2} \mo^{\eps}$.
The graphs of figure~\ref{fig1} give the perturbative expression
for $m_{F}(\mo)$ to ${\cal O}(\go^4)$, for arbitrary $N$: 
\beqa
 m_{F}(\mo)& = & m_{0}+(2N-1) \; \go^{2} \; \ggt \; \mo^{1+\eps} \nonumber \\
           &   & +(2N-1) \; g_{0}^{4} \; \ggt^{2} \; \mo^{1+2\eps} \; (1+\eps)
                 \; \Bigl\{ \,(2N-1) - \frac{\Pi(\eps)}{2} \, \Bigr\}
\nonumber \\
           &    &  + O(g_{0}^6) \hh ,
\label{ptg4}
\eeqa
where we have used the following definitions:
\beqa
\ggt & \equiv  & \frac{\Gamma(-\frac{\eps}{2})}{(4\pi)^{1+\eps /2}} \hh ,
\nonumber \\
\Pi(\eps) & \equiv & \frac{2^{-\eps}}{\sqrt{\pi}}\frac{\Gamma(1/2-\eps /2)}
 {\Gamma(-\eps /2)} \int_{0}^{1} dx \, dy \frac{[x(1-x)]^{-\eps /2}y^{-1-\eps
/2}
(1+y)}{[(1-y)^2x(1-x)+y]^{-\eps}} \nonumber \\
    &{\simeq}_{ \hspace{-15pt} \raisebox{-1.ex}{${\scriptstyle \eps
\rightarrow 0 }$ } }&
         1-\frac{3\eps}{2}+\Bigl[\frac{\pi^2}{24} -
\frac{2.9022\ldots}{2}\Bigr]\eps^2+O(\eps^3) \hh .
\eeqa
 
Now the previous expression (\ref{mf1}) 
(from now on denoted as ${\cal F}$),
\beq
{\cal F}(\mo)=\frac{\mo}{\bigl\{1-2(N-1) \, \go^2 \, \ggt \,
\mu^{-\eps}\,{\cal F}^{\eps}\bigr\}^{\frac{N-1/2}{N-1}}}
\label{F}
\eeq
reproduces eq.~(\ref{ptg4}) only to order $\go^2$ and the
leading, $1/\eps^2$ divergent terms
of order $\go^4$, while at the same time being finite, to all orders in $g^2$
fixed
for $\eps\rightarrow 0$, when the renormalized mass and coupling 
are substituted, as discussed in the previous section. 
It should be noted however that, even when restricting to the first
RG order, 
$m_{F}$ need not be identical to ${\cal F}$ : at any order in $\go^2$
there is room for a finite term, to be determined by an explicit calculation.
This follows from the fact that
\beq
\frac{\go^{2} {\cal F}^{\eps}}{\{1-2(N-1)\go^2 
\ggt {\cal F}^{\eps}\}^{1+\eps}}
\eeq
is itself, as well as all its powers, in the class of renormalization group
invariant functions of $g^2$ which are finite as $\eps\rightarrow 0$, at
fixed $g^2$.
So it is clear that $m_{F}$ can be expressed as a unique expansion in this
expression.
The above fact enables us to make compatible our variational procedure
with renormalization theory at finite $N$ as we shall see next. 
We must accordingly take into account 
the fact that equation~(\ref{F}) does not
reproduce the expansion of equation~(\ref{ptg4}) at order $\go^4$, since 
a finite term (as $\eps\rightarrow 0$) survives, as well
 as a next-to-leading
divergent term $\sim \go^4/\eps$, at that order~\footnote{
Indeed, even at order $\go ^{2}$, a finite term survives, 
if in equation~(\ref{mf1})
we replace $\Gamma(-\frac{\eps}{2})$ for example by its leading behavior,
$-2/\eps $.}.

To systematically correct
these discrepancies, while keeping at all times smooth limits
as $\eps\rightarrow 0$ for fixed renormalized parameters,
is achieved by noting that the renormalization group
invariant:
\beq
 \frac{\go^{2}\mo^{\eps}}{\{1-2(N-1)\go^{2} \ggt
\, {\cal F}^{\eps} \}^{1+\eps\frac{N-1/2}{N-1}}}
\eeq
is finite for $\eps\rightarrow 0$
and has a power series expansion
in both $\go^2$ and $g^2$.
With this invariant and all its positive powers, one will then reproduce
the whole perturbative 
expansion (in powers of $\go^2$ ) of equation~(\ref{ptg4}).

The construction of our generalized Ansatz therefore goes as follows.
We consider 
\beq
{\cal F}(\mo)=\frac{\mo}{\bigl\{1-2(N-1)\go^2\ggt 
{(\frac{{\cal F}}{\mu})}^{\eps}\bigr\}^{A}}
\hh ,
\label{Fa}
\eeq
entering the following renormalization group invariant
expression for $m_F$, 
\beqa
m_{F}(\mo) & = & \frac{\mo}{\bigl\{1-2(N-1)\go^2\ggt
{(\frac{{\cal F}}{\mu})}^{\eps}\bigr\}^{B}} \;
\Bigl[ \; 1 + \frac{ \lambda \; \go^2 \mo^{\eps} }
{\bigl\{1-2(N-1)\go^2\ggt {(\frac{{\cal F}}{\mu})}^{\eps}
\bigr\}^{D}}  + \nonumber \\
& & \frac{ \rho  \;  \go^4 \mo^{2 \eps} }
{\bigl\{1-2(N-1)\go^2\ggt {(\frac{{\cal F}}
{\mu})}^{\eps}\bigr\}^{2 D}} +\ldots \Bigr]
\hh ;
\label{mf2}
\eeqa
 
where the constants $A$, $B$ and $D$ 
are to be determined by {\em requiring}
equation~(\ref{mf2}) to be finite as $\eps\rightarrow 0$ when expressed in
terms of $m$ and $g^2$. $A$, $B$ and $D$ are therefore obtained 
as an expansion
in powers of $\eps$, where the different orders of the RG coefficients
enter at the different order of $\eps$. 
The coefficient $\lambda$, $\rho$, $\cdots$
are then given by matching the perturbative expression for $m_F$,
equation~(\ref{ptg4}). This is explained in more details in Appendix~B.

With expression (\ref{mf2}) and a straightforward generalization 
of the steps decribed in Appendix A, 
one can reach infinite order 
in the variational expansion
in a very similar way as the one leading to 
equation~(\ref{MF}), (\ref{MFF}), and obtains the following generalized
version
of equation~(\ref{MFF}): 
\beqa
m_{F}(m'') & = &
\ds { \Bigl(\frac{\rm e}{2}\Bigr)^{-\frac{1}{2(N-1)}}
\;\Lam \;\times } \nonumber \\
 & & \ds {\frac{m''}{2i\pi}\oint\frac{du{\rm e}^{u}}
{f_{2}^{\frac{N-1/2}{N-1}}(u)}}
\Bigl\{ 1+\frac{{\cal M}_{1}}{f_{2}}+\frac{{\cal M}_{2}}{f_{2}^{2}} +\cdots
\Bigr\} \hh ,
\label{mresult}
\eeqa
where now~\footnote{Note the difference between the 
coefficient of the $\ln f_2$
term in (\ref{mcoef}) and the power of $f_2$ in (\ref{mresult}). The previous
equality of those quantities, in eq.(\ref{MF}) (and the consequent
simple pole behavior for $m'' \to 0$) was an accident of the first RG order.}
\beqa
f_2(u) & = & \ln(m''u)-\frac{N}{N-1}\ln f_2(u) \hh , \nonumber \\
{\cal M}_{1} & = & \frac{3 (N-1/2)}{4 (N-1)^2} \hh , \nonumber \\
{\cal M}_{2} & = & - \frac{(N-1/2)(36N^2-62N+17)}{64(N-1)^4}\nonumber
\\
 & & -\frac{N-1/2}{2(N-1)^2}\bigl(-0.725551\ldots + \frac{\pi^2}{48}\bigr) 
\hh ,
\label{mcoef}
\eeqa
and in the overall factor in (\ref{mresult}) we again identifed the usual
renormalization group invariant scale at this order,
$$\Lambda_{\overline{MS}}=\overline{\mu}\;
{\rm exp}\Bigl[-\frac{\pi}{(N-1)g^2}\Bigr]\; \Bigl [
\frac{(N-1)g^2}{2\pi}\Bigr ]^{\frac{1}{2(N-1)}}
 \Bigl(1-\frac{g^2}{2\pi}\Bigr)
^{-\frac{1}{2(N-1)}}\; ,
$$
up to an overall 
factor independent of $g^2$ in (\ref{mresult}).
 
We now discuss expression (\ref{mresult}), 
which is the essential result of this paper.
One can analyze the function $m_{F}(m'')$ in standard ways.
One finds that for $m''\rightarrow\infty$, its behavior
is accurately  described by the first term of equation~(\ref{mresult})
\beq
m_{F}(m'')\sim
\;\Lambda_{\overline{MS}} \ds { \Bigl(\frac{\rm
e}{2}\Bigr)^{-\frac{1}{2(N-1)}} }
\frac{m''}{\bigl(\ln{m''}\bigr)^{\frac{N-1/2}{N-1}}} \hh ,
\label{asymp}
\eeq
a typical renormalization group result, and that the
${\cal M}_{n} / f_{2}^{n}$ correction contributes a corrective term
of order $(\ln{m''})^{-n}$ to this, as well as $\ln \ln m''$ terms.
Hence, by going to higher and higher orders of ordinary perturbation
theory, one determines the function $m_{F}(m'')$ with an increasing
accuracy for $m''$ large enough.
 
The same phenomenon had been noticed for the anharmonic oscillator
case in references~\cite{BGN1,BGN2}.
Just as in that case, the hope is then that already with a low
order calculation, one may reach an accurate value for, say,
the physical mass gap by taking an extremum of $m_{F}(m'')$,
the best one being presumably the one closest to $m''=0$.
For the $N=\infty$ case, we have seen that all the coefficients
${\cal M}_{n}$ are zero, and the extremum at $m''=0$
is the exact answer. For finite $N$, the numerical convergence
of the approximation can only be decided by an explicit calculation.
 
Notice that, when the order of the perturbative expansion increases,
one  can deduce from the behavior of $f(u)$ as $u$ goes to zero,
that $m_{F}(m'')$ becomes more and more singular as $m''\rightarrow 0$.
This same fact had been noticed on the case of the anharmonic oscillator.
There, this singular behavior at $m''=0$ did not prevent the analogous
function from {\em  converging} to the right answer for a wider and wider
range of the variational parameter when the order of the expansion increases.
Thus, one could obtain very accurate results, in particular
when resummation techniques are used to improve the behavior of the
integrand of equation~(\ref{mresult}) for $u\rightarrow 0$. This will
be discussed below.
\subsection{Vacuum energy density} 
A similar treatment can be applied to the vacuum energy density
$E_{0}^{(B)}$, with an appropriate modification to account for the
structure of its divergences.
 
Indeed, in contrast with $m_{F}$, the vacuum energy, in the massive theory,
 is not multiplicatively renormalizable. Rather, 
$\frac{\partial^3 E_{0}^{(B)}}
{\partial\mo^3}$ is, and consequently $\frac{\partial^3 E_{0}^{(B)}}
{\partial m^3}$ is finite.
Combining this fact with dimensional arguments, it follows that
all infinities of  $E_{0}^{(B)}$ which persist, at a given
order of the perturbative expansion (after coupling constant,
mass, and wave function renormalization), are proportional to
$\frac{\mo^2}{2\go^2}$. Hence, by subtracting from $E_{0}^{(B)}$ the quantity
$\frac{\mo^2}{2\go^2}h(\eps)$
with a suitably chosen function $h(\eps)$,  $E_{0}^{(B)}$ can be
made finite for $\eps\rightarrow 0$ at fixed renormalized coupling
constant $g$ and mass $m$. 
It is crucial to remark that such a subtracted counterterm gives vanishing
contribution to our procedure {\it for any} $ \eps$ as soon as the order of
the variational-perturbative expansion is larger than one\footnote{
More precisely
the subtraction procedure does not give
any new contributions
beyond the generic form in (\ref{eresult}) below, but 
actually the specific values of the perturbative coefficients, 
${\cal E}_i$
in (\ref{eresult}), consistently include a dependence on 
the subtracted terms.}.
Expanding now the regular function  $h(\eps)$ as
\beq
h(\eps)=\sum_{n \geq 0} h_{n}\eps^n \hh ,
\eeq
one can uniquely determine the coefficient of the $\eps^n$ term
by a direct perturbative calculation to $n+1$ loops. In contrast
with the mass case, the fact that
the $(n+1)$ loop information is needed to determine unambiguously
the $1/f^n$ perturbative corrections in our Ansatz, is of course
a reminiscence of the original ambiguity due to the above mentioned 
additional
divergences of the vacuum energy~\footnote{Actually however, only the 
value of the simple pole in $\eps$ at order $n+1$ is needed.}. 
 
Once this is done, the finite terms that are left over can be
treated in the same way as $m_{F}$ was treated.
We skip the details and only give the final answer, as
a function of the variational parameter $m''$, at $\eps=0$
\beqa
\label{eresult}
E_{0}(m'') & = & 2 (N-1)\; \Bigl(\frac{\rm e}{2}\Bigr)^{-\frac{1}{(N-1)}} \;
\Lambda^2_{\overline{MS}} \\ \nonumber
 & & \ds \times {\frac{{m''}^2}{2i\pi}\oint\frac{udu{\rm e}^{u}}{f_{2}^
{\frac{N}{N-1}}(u)}}
\Bigl\{ 1+\frac{{\cal E}_{1}}{f_2}+\frac{{\cal E}_{2}}{{f_2}^{2}}+
\ldots \Bigr\} \; ,
\eeqa
where
\beqa
{\cal E}_{1} & = & \frac{N (N^{2}+N/2-3/4)}{2 (N-1/2)(N-1)^2} \hh , \nonumber
\\
{\cal E}_{2} & = & - \frac{N (N-1/2)(N^{2}- 6 N + 7/2)}{8 (N-2/3) 
(N-1)^4} \hh
.
\label{ecoef}
\eeqa
 
\subsection{Refinements}\label{refin}
 
While our main result formulas (\ref{mresult}) and (\ref{eresult}) 
contain in their derivation all the conceptual ingredients
of the method, they leave room for some refinements, which are necessary
before they can be of practical use for optimization.
 
The first one comes from realizing that, even at a given RG order,
 our different Ans\"atze are 
not unique, since there are infinitely many ways of introducing the
purely perturbative terms, $1/f^n$ corrections. (Moreover
beyond the first RG order even the
resummation of the correct two-loop RG behaviour is not
uniquely fixed by the function ${\cal F}$ as defined 
in eq.~(\ref{mcoef}), as will be discussed in the next section). 
To take into account this freedom, let us introduce an
arbitrary parameter
$a$, from substituting 
$\bar \mu \to a\;\bar\mu$ in the different relevant expressions. 
This constant, which enlarges the class of our
\RG\ invariant finite guesses for $m_F$ , cannot be ruled out {\em a priori},
and parameterizes a renormalization scheme (RS) dependence.
Of course, if one were able to work to all orders, the final answer
would be independent on $a$. However, since we are truncating the 
expansion in
equation~(\ref{mf2}), the choice of this scaling constant $a$ turns out to be
important in order to obtain a reliable numerical estimation for $m_F$.
In the present model, the value of $a$ which ensures a rapid
convergence of the expansion in equation~(\ref{mf2})
is expected to be  $a=1+O(1/N)$. We have however no {\it a priori} idea
 of the value to chose in more complicated
theories as  QCD. Hence we shall see in the
following how to cope with the existence of the parameter $a$,
intrinsic to renormalizable theories,
 in order to avoid too much additional arbitrariness in our results.
 
Applying now the variational procedure, we finally obtain a last version of
equation~(\ref{mresult}), expressed in terms of the usual \RG\ invariant
parameter $\Lambda_{\overline{MS}}$ ,
\beq
\frac{m_{F}}{\Lambda_{\overline{MS}}}  =
\Bigl(\frac{\rm e}{2}\Bigr)^{-\frac{1}{2(N-1)}} \; \ds 
{\frac{m''\:a}{2i\pi} \;
\oint\frac{du{\rm e}^{u}}{f_{2}^{\frac{N-1/2}{N-1}}(u)}}
\Bigl\{ 1+\frac{{\cal M}_{1}(a)}{f_{2}}+\frac{{\cal
M}_{2}(a)}{f_{2}^{2}}+\ldots \Bigr\} \hh ,
\label{mgap2c}
\eeq
where the recursively defined function $f_2$ is given by
\beq
f_2 + \frac{N}{N-1} \ln f_2 = \ln \Bigl(\frac{m'' u}{a}\Bigr) \hh .
\label{recur2}
\eeq
The coefficients in the expansion are now given by
\beqa
{\cal M}_{1}(a) & = & {\cal M}_{1}(1)-\frac{(N-1/2)}{(N-1)} \ln a  \hh ,
\nonumber \\
{\cal M}_{2}(a) & = & {\cal M}_{2}(1)+\frac{(N-3/4)(N-1/2)}{(N-1)^2} \ln^2 a
+ \nonumber \\
&  & \frac{(N-1/2)(N^{2}- 5/2 N + 9/8)}{(N-1)^3} \ln a \hh ,
\eeqa
where ${\cal M}_{1}(1)$ and ${\cal M}_{2}(1)$ are defined in
equation~(\ref{mcoef}).
 
Similarly, we obtain for the energy
\beq
\frac{E_{0}}{2 (N-1) \Lambda_{\overline{MS}}^2}   =
\Bigl( \frac{{\rm e}}{2} \Bigr) ^{-\frac{1}{(N-1)}} \;
\ds {\frac{(m'')^2\:a^2}{2i\pi}
\oint\frac{udu{\rm e}^{u}}{f_{2}^{\frac{N}{N-1}}(u)}}
\Bigl\{ 1+\frac{{\cal E}_{1}(a)}{f_{2}}+\frac{{\cal
E}_{2}(a)}{f_{2}^{2}}+\ldots \Bigr\}
\label{energ2c}
\eeq
with
\beqa
{\cal E}_{1}(a) & = & {\cal E}_{1}(1)-\frac{N}{(N-1)}\ln a \hh , \nonumber
\\
{\cal E}_{2}(a) & = &  {\cal E}_{2}(1) + \frac{N(N-1/2)}{(N-1)^2} \ln^2 a -
\frac{3 N (N-1/2)}{2(N-1)^2} \ln a \hh ,
\eeqa
${\cal E}_{1}(1)$ and ${\cal E}_{2}(1)$ being given in equation~(\ref{ecoef}).
 
We are now asked to find a stationary value for these expressions in term
of $m''$ :
\beq
\frac{\partial m_{F}}{\partial m''}=0 \hh \mbox{ and }
\hh \frac{\partial E_{0}}{\partial m''}=0 \hh .
\eeq
 
In order to get a feeling of what we have achieved,
we have plotted in figure~\ref{fig3} the values of the mass gap
$m_{F}/\Lambda_{\overline{MS}}$ versus the variational parameter $m''$,
for different values of the scaling constant $a$ ranging from $a=1.1$ to
$a=1.2$. Similarly, we have plotted in figure~\ref{fig4}  the values of
the vacuum energy $E_{0}/{\Lambda_{\overline{MS}}}^2$ versus  $m''$,
for $a$ ranging from $a=0.9$ to $a=1.1$. In both cases, we have taken a
 reasonably small number of ``flavors", $N=3$.
We have limited the graph to the vicinity of the origin in $m"$
since it is the region of interest (for large $m"$
 the curves are following the expected perturbative behavior). 
The dashed
horizontals represent the known exact values for the
mass~\footnote{In these figures, we have used
an expansion  for $m''\rightarrow 0$ in order to produce
the curves in a reasonable computing time. The corresponding slight change in
 the shapes of the displayed curves is irrelevant
for the discussion.}~\cite{FNW}, obtained from the Bethe Ansatz:
\beq
\frac{m_F}{\Lambda_{\overline{MS}}}=
\frac{(4\rm e)^{\frac{1}{2(N-1)}}}{\Gamma\Bigl(1-\frac{1}{2(N-1)}\Bigr)}\hh ,
\label{mexact}
\eeq
and for the vacuum energy\footnote{We are grateful to
Prof. Al. B. Zamolodchikov for communicating his result to us
before publication.}~\cite{ZAM}
\beq
E_0=-\frac{m_F^2}{8} \cot \frac{\pi}{2(N-1)} \hh .
\label{Eexact}
\eeq
 
 Many comments are now in order:
 
First of all, we have succeeded in producing finite values for the mass gap
$m_{F}$
and the vacuum energy $E_0$
at 2-loops order, including a variational mass $m''$, in a way consistent
with the \RG\ . The extension of the procedure to higher orders
in perturbation is only a matter of computational effort. Moreover, the
method can be applied, at least in principle, to some of the
order parameters governing the chiral symmetry breakdown
in the realistic case of QCD, as we shall see in a subsequent paper. 
 
Secondly, these expressions, equations~(\ref{mgap2c}) and~(\ref{energ2c}),
 exhibit stationary values with respect to the parameter $m''$, as
can be seen on figures~\ref{fig3} and \ref{fig4} . We can then proceed
and look at these stationary values.
 
We immediately meet an obvious problem: due to the large dispersion
of the curves when the scaling constant $a$ is varied, it seems
difficult to predict a reliable numerical value for the
 ratios $m_{F}/\Lambda_{\overline{MS}}$ and
$E_{0}/\Lambda^2_{\overline{MS}}$. Actually, the stationary values for the
ratio $m_{F}/\Lambda_{\overline{MS}}$ are seen to lie in the range~1 to 3
when $a$ is varied from 1.1 to 1.2, and one cannot hope to extract from
that more than a vague estimate of the mass gap.
 
However, one can see in figure~\ref{fig3} that the flattest curve,
whose stability plateau gets closest to $m''=0$, lies quite near to the
expected exact value. Moreover, the divergence of the curves near $m''=0$,
where we would expect to obtain the exact answer, as in the
$N\rightarrow\infty$ case, is only due to the breakdown of
perturbation theory as $m''\rightarrow 0$. This, and previous experience
with the anharmonic oscillator case~\cite{BGN1}, suggest us to try an
extrapolation towards $m''= 0$ in order to improve the numerical
precision of our results. This will be done now.
\subsection{Optimization results} 
The strategy is to turn the expansion in powers of $1/f_2$
of equation~(\ref{mgap2c}) into an expansion
around $m''=\infty$, whose first term appears in 
equation~(\ref{asymp}), later
to be extrapolated by some variant of Pad\'e techniques to $m''=0$.
 This however produces an expansion containing $\ln \ln  \ldots \ln m''$ ,
which is not suitable for an extrapolation towards $m''= 0$.
In fact, equation~(\ref{recur2}) suggests that instead of $m''$ one
should use the variable $\eta$ defined by
\beq
\eta + \frac{N}{N-1} \ln \eta = \ln \Bigl(\frac{m''}{a}\Bigr) \hh .
\label{defeta}
\eeq
 
In terms of $\eta$, the $\ln \ln \ldots \ln m''$ disentangle, and
expression~(\ref{mgap2c}) now admits a very nice asymptotic expansion
for $\eta \rightarrow \infty$, which is readily seen 
to give\footnote{In (\ref{eta}) the coefficients of
$\eta^{-n}$ involve integrals of the type
$I(p) = \frac{1}{2i\pi} \oint dy \: e^y \ln^p y$, which are easily evaluated
as a (finite) expansion of real integrals expressable 
in terms of Euler Gamma functions.}
 
\beqa
\frac{m_{F}}{\Lambda_{\overline{MS}}} & = &
\Bigl(\frac{\rm e}{2}\Bigr)^{-\frac{1}{2(N-1)}} \;\frac{N-1/2}{N-1} \;
 a {\rm e}^{\eta}
{\eta}^{-\frac{N-3/2}{N-1}} \;
\Bigl\{ \; 1 \; + \;
\frac{1}{\eta} \; \Bigl( \; 2 \; \frac{N-3/2}{N-1/2} \; {\cal M}_{1}(a) \; 
\; \nonumber \\
 & + & \frac{(2N-3/2)}{N-1} \; \gamma_{E} \; - \; 
\frac{N}{N-1} \;  \Bigr) + \;
 O(1/\eta^2) \; \Bigr\} \hh .
\label{eta}
\eeqa
 
Taking now the derivative of the logarithm of this expression yields a power
series in $1/\eta$ , which is suitable for a Pad\'{e} approximant 
analysis:
\beqa
\frac{\partial \ln \left( \frac{m_{F}}{\Lambda_{\overline{MS}}} \right)
}{\partial \eta } & \simeq &
1 \; - \; \frac{1}{\eta} \, \frac{N-3/2}{N-1} \; - \;
\frac{1}{\eta^2} \, \Bigl( \; 2 \; \frac{N-3/2}{N-1/2} \; 
{\cal M}_{1}(a) \; +
\; \nonumber \\
 & & \frac{(2N-3/2)}{N-1} \; \gamma_{E} \; - \; \frac{N}{N-1} \;  
\Bigr) \; +
\;\ldots \hh .
\label{pade}
\eeqa
 
From the large-$N$ case studied in a preceding paper, we can make
the guess that, were it not for the fact that perturbation theory breaks
down at $m'' = 0$, $m_{F}(m'')$ would approach the exact mass gap as
\beq
m_{F}(m'') \hspace{10pt} {\simeq}_{ \hspace{-20pt}
\raisebox{-1.ex}{${\scriptstyle m'' \rightarrow 0 }$ } }
m_{F}(0) \; + \;  O ({\rm e}^{-{\rm Const}/m''} ) \hh .
\eeq
 
Therefore we can try a Pad\'{e} approximant of the form
\beq
\frac{\partial \ln \left( \frac{m_{F}}{\Lambda_{\overline{MS}}} \right)
}{\partial \eta } \simeq
{\rm e}^{{C_1}/{\eta}} \; \frac{1 + {C_2}/{\eta} + {C_4}/{\eta^2} + \ldots }
{1 + {C_3}/{\eta} + {C_5}/{\eta^2} + \ldots } \hh ,
\label{padeansatz}
\eeq
which ensures an exponential convergence of $m_F$ near $m''=0$,
{\em provided} that $C_1 < 0$.
However this is by no way limitative, in particular for {\em finite}
$N$ there are no guarantees that the exponential behavior persists, 
therefore we shall also use simple Pad\'e approximants (i.e
similar to (\ref{padeansatz}) 
without exponential factor), and compare 
different Pad\'e results. 
For any Pad\'e approximant type we
can now determine the coefficients 
 \{ $C_1$ , $C_2$ , $C_3$ \ldots \}  by
matching the large-$\eta$ expansion~(\ref{pade}), and integrating
equation~(\ref{padeansatz}) from $\eta \rightarrow +\infty$,
where $m_F$ is given by equation~(\ref{pade}), to $\eta = 0$, will produce an
extrapolated
value  at $m''=0$ for the mass gap
$m_F/\Lambda_{\overline{MS}}$~\footnote{Note that
the re-integration of $m_F/\Lam$ 
is unique, as the integration constant is fixed by the asymptotic
$m'' \to \infty$ behavior.}.

Very likeky, it should exist an optimal Pad\'e order, which may however
vary as $N$ varies, and is difficult to guess a priori. 
One may naively assume at first that the best Pad\'e
approximant should be of the same order of the known perturbative terms.
Actually the situation is slighlty more complicated, 
since the RG coefficients, which is a genuine
information independent of the purely perturbative terms, also enter
Pad\'e approximant 
expansions. Moreover, the larger $N$ is, the more perturbative
information may be considered 
available, since {\it all} the purely 
perturbative
coefficents ${\cal O}(1/f^n)$ in $m_F$ 
are vanishing for $N \to \infty $ (although we do not know 
{\em how fast} these are vanishing as functions of N). 
Hence, one can test the convergence
properties of the increasing order  
Pad\'e approximants~(\ref{padeansatz}) in the large-$N$ case, and
infer from it the accuracy of the lower Pad\'es at finite $N$.
Without much loose of generality in our numerical study 
we compare Pad\'e approximants
of different orders, from (1,1) to (2,3).  
 
The results of the procedure for $N \to \infty$ are presented in
table~\ref{tabl1}, where we have given for different values of the scaling
constant
$a$, the value of the extrapolated ratio, 
using a Pad\'{e} 1-1, a Pad\'e 2-2 or
a Pad\'e 3-3 approximant\footnote{The blank entries 
correspond to values of $a$
where the coefficient  $C_1$ is positive. $C_1$ is
a complicated function of $a$, and
there may be more than one range of values of $a$
where it is negative, which happens
for the first time for the 3-3 Pad\'e.}.
 The exact value is  $m_F/\Lambda_{\overline{MS}}=1$.
We see that the stationary (in $a$) values of the successive Pad\'es do
indeed converge, albeit
rather slowly, as the order increases, and with what 
may be considered relatively
large errors (20\% for the 1-1 Pad\'e). This indicates that the extrapolation
procedure is probably far from optimal, and could be improved by 
using information
on the asymptotic behavior of the large-$\eta$ expansion for example. 
 
\begin{table}[htb]
\centering
\begin{tabular}{||l||l|l|l||}  \hline
$a$ & Pad\'{e} 1-1 & Pad\'{e} 2-2 & Pad\'e 3-3 \\ \hline
0.1  &          &  0.9643  &            \\ \hline
0.18  & 1.2115  &  0.9913  &  1.0031    \\ \hline
0.2  & 1.2030   &  0.9965  &  1.0044    \\ \hline
0.25  & 1.2041  &  1.0080  &  1.0093     \\ \hline
0.5  & 1.3115   &  1.0463  &             \\ \hline
0.7  & 1.4379   &  1.0623  &  0.9526      \\ \hline
0.8  & 1.5105   &  1.0668  &  0.9594      \\ \hline
0.9  & 1.5896   &  1.0696  &  0.9602      \\ \hline
1.0  & 1.6760   &  1.07095 &  0.9582     \\ \hline
1.1  & 1.7711   &  1.07096 &  0.9544     \\ \hline
1.2  & 1.8768   &  1.0698  &  0.9491     \\ \hline
1.3  & 2.0025   &  1.0676  &  0.9422      \\ \hline
1.5  & 2.2636   &  1.0602  &  0.9231       \\ \hline
opt. & 1.203    &  1.071   &   0.9602     \\ \hline
      &         &          &   1.002      \\ \hline
error &  +20\%   &  +7\%     &  -4\%,\,\, +0.2\%       \\ \hline
\end{tabular}
\caption{extrapolation of $m_F/\Lambda_{\overline{MS}}$ towards $m''=0$ using
Pad\'{e} approximants in the case $N=\infty$. }
\label{tabl1}
\end{table}
 
For finite $N$, different Pad\'e approximants 
give quite different answers,
however it turns out that on the average the results are not worst than 
in the large N case. 
Results for the extrapolation of $m_F/\Lambda_{\overline{MS}}$
 to $m^"=0$ are presented in table~\ref{tabl2}
for different values of $N$ and different Pad\'e types,
where we show the optimal values and the
 exact values from reference~\cite{FNW}. In addition we illustrate 
in figure 5
a comparison of the a dependence for the different
Pad\'e types, for 
the lowest value of $N=2$, 
the behavior for other values being very similar.

 We note the great stability as a function of $a$ when $a$ is varied in
a wide range, in contrast with the dispersion of the curves in 
figure~\ref{fig3}.
We see that in some cases 
the optimal values are quite remarkably close to the exact ones, even for 
very low values of $N$. This is encouraging especially in view
of the fact that the (only known)
first two non-RG perturbative corrections were used.
Actually however, the extremely good agreement for 
specific 
Pad\'e types, for instance  at $N=2$ and $N=3$,
 should be attributed to numerical occasion.
What is certainly more significant is the {\em average}
error over different values of $N$, for
a given Pad\'e order and type. 
The overall best result for arbitrary $N$ 
are obtained with the simple Pad\'e (2,3) and the exponential 
Pad\'e (2,2) which
we note are both using expansion to the fifth order in $1/\eta$ in 
(\ref{pade}).  
This is most probably due to the fact that, at the order where we
are working, for a fixed $N$ and $a$,  
expression (\ref{mgap2c}) actually depends on five
{\em independent} parameters: three perturbative terms, namely
the ``zero order" overall coefficient, ${\cal M}_1$, and
${\cal M}_2$; {\em plus}
 two independent combinations of RG coefficients, namely
the coefficient of $\ln f_2$ in (\ref{mcoef}) and the power of $f_2$ in
(\ref{recur2}). However it is not
excluded that higher Pad\'e orders would give even better results.
Due to the rather complicated dependence of a given Pad\'e upon the
above parameters, we have not tried to systematically study this issue.  
It would certainly be of interest to further optimize the choice of
the Pad\'e approximants,  
but this  
 would go beyond the scope of the present paper.
At any rate, 
given the very different Pad\'e approximant types confronted here,
we can be confident that we
have found a useful convergent variational scheme to compute non-perturbative
quantities.
 
\begin{table}[htb] \centering
\begin{tabular}{||l||l|l|l|l||}  \hline
 & $N=2$   &  $N=3$  &  $N=5$       &   $N=8$       \\ \hline
exact result & 1.8604   &  1.4819    &    1.2367   &    1.133   \\ \hline
Pad\'e type & & & & \\ \hline
$p_{1,1}(u) \times e^{(c1/u)}$ &  &  & &               \\
opt. result (error)
& 2.742 (32\%)&  1.83375 (19\%)&  1.48725 (17\%) & 1.3554 (16\%)
\\ \hline
$p_{1,2}(u)$ & & & & \\
opt. result (error)
& 1.9278 (3.5\%) & 1.3079 (11.7\%) & 1.08105 (12.6\%) & 0.9961 (12\%)
\\ \hline
$p_{2,2}(u) \times e^{(c1/u)}$ & & & &    \\
opt. result (error)
& 1.758 (5.5\%) & 1.47498 (0.46\%) & 1.2843 (3.7\%) & 1.19626 (5.3\%)
\\ \hline
$p_{2,3}(u) $ & & & & \\
opt. result (error)
& 1.6206 (12.8\%) & 1.3456 (9.2\%) & 1.1917 (3.6\%) & 1.1205 (1.1\%)
\\ \hline
$p^\prime_{2,3}(u)$ & & & & \\
opt. result (error)
& 1.8749 (0.77\%) & 1.4864 (0.3\%) & 1.2654 (2.3\%)& 1.1628 (2.6\%)
\\ \hline
\end{tabular}
\caption{Extrapolation of $m_F/\Lambda_{\overline{MS}}$ towards $m''=0$
using different Pad\'e approximants for finite $N$. The last two lines
provide in addition a comparison of different RS
for a same Pad\'e approximant order: accordingly
the results given in the last line were obtained using
the alternative Ansatz form (\ref{mgap2d}) derived in section 4. }
\label{tabl2}
\end{table}
 
A particular feature of renormalizable  theories might however
spoil this optimism: it is the dependence of perturbative computations on the
renormalization scheme (RS). As is well known, 
apart from the arbitrariness of the
renormalization scale, parametrized with $a$, 
already at the second RG order there is more arbitrariness
in the renormalization prescriptions. 
Typically, only the first two
perturbative coefficients of the beta function and the first 
coefficient of the
anomalous mass dimension 
are scheme independent, so that higher coefficients can be
set to arbitrary values by perturbative redefinitions 
of the coupling constant
and finite renormalizations. In our case, this translates
into the fact that in principle the higher
 coefficients ${\cal M}_{i}$ for $i$
larger than one could be set to arbitrary values with an {\it ad hoc}
renormalization scheme. This problem is well known in perturbative QCD.
The only assurance is that any {\em reasonable} renormalization scheme should
give similar results at a given order, and that the accuracy of the results
should improve as the order increases. 
What reasonable means exactly is unknown,
but dimensional regularization with minimal subtraction is generally believed to
be such a reasonable scheme. To test convergence and accuracy, we do not 
have higher
order calculations at our disposal, but we can use the other method: do the
same computation at the same order with a different renormalization
scheme (RS) which can be considered {\it a priori} as reasonable, and
compare the results. This is what we now do. 
\section {Alternative Scheme Ansatz}\label{alternative}
\setcounter{equation}{0}
The \RG\ invariant expression (\ref{mf2}) has been 
reconciled in section 3 with a variational mass expansion, 
using \RG\ invariance
properties of appropriate combinations of {\it bare} parameters,
$m_0$, $g_0$ like in equation~(\ref{Fa}), and the resummation
properties of the $x$ series, resulting in the integral result
(\ref{mresult}). However, once this construction is understood, and 
a finite, renormalized mass gap (\ref{mgap2c}) emerges,
one can alternatively try to construct an 
Ansatz starting directly from {\it renormalized}
quantities\footnote{This procedure
will also be treated in more details in the QCD context~\cite{
AGKN}.}. This will turn out to provide a more transparent intepretation
of the different quantities involved in (\ref{mresult})--(\ref{mgap2c}),
which also turns out to be more appropriate for a generalization to other
theories, like QCD typically. 

Let us thus start from the \RG\ invariance properties of the renormalized
mass and coupling constant, and examine what kind of non-perturbative
information it may contain. Integrating the
\RG\ equations for the running
mass $m(\bar \mu)$ in terms of the running coupling constant,
$g(\bar \mu)$, to two-loop \RG\ order exactly\footnote{{\it i.e.} keeping
the exact dependence on $g$ as given by the \RG\, but of course
the \RG\ functions themselves being truncated at a given order.},
with the ``fixed point" boundary condition
$$
M^{RG}_F \equiv m(M^{RG}_F),
$$
one obtains after some algebra the expression
 
\beq
M^{RG}_F =  m(\bar \mu) \;\;
\ds{f^{-\frac{ \gamma_0}{2b_0}}\;\; \Bigl[\frac{ 1 +\frac{b_1}{b_0} 
g^2(\bar\mu) f^{-1}}{ 1+\frac{b_1}{b_0}g^2(\bar \mu)} 
\Bigr]^{ -\frac{\gamma_1}{2 b_1}
+\frac{\gamma_0}{2 b_0}   } }\;
\label{MRG}
\eeq
where
$f \equiv g^2(\bar \mu)/g^2(M^{RG}_F)$ satisfies the recursive relation
\beq
f = \ds{ 1 +2b_0 g^2(\bar \mu) \ln \frac{M^{RG}_F}{\bar \mu }
+\frac{b_1}{b_0} g^2(\bar \mu)
\ln \Bigl[\frac{ 1 +\frac{b_1}{b_0} g^2(\bar \mu) f^{-1}}{
 1 +\frac{b_1}{b_0} g^2(\bar \mu) }\;f\;\Bigr] }\; ;
\label{frenrec}
\eeq
and the \RG\ coefficients $b_i$ and $\gamma_i$ are given in Appendix~B.
$M^{RG}_F$ designates the part of the mass which is entirely
determined from \RG\ properties, {\it i.e.} it does not
include the purely perturbative non-logarithmic finite parts.
The latter are consistently 
included as follows:
\beq
M_F \equiv M^{RG}_F \;\Bigl(1 +m_{11}\frac{g^2(\bar \mu)}{f}
+m_{22}\frac{g^4(\bar \mu)}{f^2}+\dots\;\Bigr)
\label{MFall}
\eeq
with $m_{11}$ and $m_{22}$ simply given by the non-logarithmic
parts of the perturbative expression for the renormalized mass,
which is evidently obtained from equation~(\ref{ptg4})
by substituting $m_0 =Z_m m$, $g^2_0 = (\bar \mu)^{-\eps} Z_g g^2$.
Using $Z_m$, $Z_g$ at second order, one finds
(in the $\overline{MS}$ scheme)
\beqa
\label{mpertfinite}
 M^{pert}_F  & =  &  m(\bar \mu)
\Bigl( 1 - g^2(\bar \mu)\frac{N-1/2}{\pi} \ln \frac{m(\bar
\mu)}{ \bar \mu}  \\ \nonumber
& &  +g^4(\bar \mu) \Bigl[\;
\frac{(N-1/2)(N-3/4)}{\pi^2} \ln^2\frac{m(\bar \mu)}{
\bar \mu}\\ \nonumber
& & +\frac{(N-1/2)(N-1/4))}{\pi^2} \ln \frac{m(\bar \mu)}{
\bar \mu} \\ \nonumber
& &  +\frac{N-1/2}{\pi^2}\bigl(0.737775-\frac{\pi^2}{96}\bigr)\;\Bigr]\; 
\Bigr) 
\hh
\eeqa
from which we immediately obtain 
\beq
m_{11} = 0 \;,\;\;\;\; m_{22}
=\frac{N-1/2}{\pi^2}\bigl(0.737775-\frac{\pi^2}{96}\bigr)\;.
\eeq
One can check that  expansion~(\ref{MFall}) then
reproduces the pertubative result~(\ref{mpertfinite}), and in fact
generates correctly, by construction,
the leading and next-to-leading logarithms, $g^{2n}\ln^n [m(\bar \mu)
/\bar \mu]$ and $g^{2n}\ln^{n-1} [m(\bar \mu)
/\bar \mu]$ respectively, to all orders.

Next, we introduce the variational mass expansion
in analogy with 
the procedure derived in section 2, by simply {\em assuming} 
the result to be given from performing the substitution 
\beq
m(\bar \mu) \to m(\bar \mu)\; u
\label{mbarsub}
\eeq
everywhere in expressions (\ref{MRG}), (\ref{frenrec}), (\ref{MFall}),
and integrating the resulting expressions along an appropriate
contour in the complex $u$ plane, with the weight $du u^{-1}\;e^{u}$,
similarly to e.g expression (\ref{mresult}).
This leads to the final formula,
\beq
{ m_F (a)\over {\Lambda_{\overline{MS}}}}
 = {2^{-C} a\: m''' \over{2 i \pi}} \oint_{\cal C} du {e^u
\over{f^{\prime A}_2 [C + f^\prime_2]^B}} {\Bigl(1 +{{\cal M}^\prime_1(a)
\over{f^\prime_2}}
+{{\cal M}^\prime_2(a)\over{f^{\prime 2}_2}} +\ldots \Bigr)},
\label{varM2}
\eeq
where we defined
$f^\prime_2 \equiv f/(2b_0g^2(\bar \mu)) = [2b_0g^2(M^{RG}_F)\:]^{-1}$
which satisfies
\beq
f^\prime_2(u) = \ln [m''' u] -A \ln f^\prime_2 -
(B-C)
\ln [C+f^\prime_2] \; .
\label{fprime2}
\eeq
In eqs. (\ref{varM2}) and (\ref{fprime2}), A, B, C are defined as
\beqa
A &=& {\gamma_1\over{2 b_1}}=\frac{N-1/2}{2(N-1)}\; , \\ \nonumber
B &=& {\gamma_0\over{2 b_0}}
-{\gamma_1\over{2 b_1}}=\frac{N-1/2}{2(N-1)}\; ,  \\  \nonumber
 C &=& {b_1\over{2b^2_0}} = -\frac{1}{2(N-1)}\; .
\hh
\eeqa
and
\beq
\label{mtierce}
m'''\equiv  \ds{\frac{m(\bar \mu)}{ \Lambda_{\overline{MS}} \;
2^{-C}\;[2b_0g^2(\bar \mu)]^{A+B}
\;[1+\frac{b_1}{b_0}g^2(\bar \mu)]^{-B}}}
\; .
\hh
\eeq
In equation~(\ref{varM2}), we also included the RS dependence through the
parameter $a$, in a way similar
to (\ref{mgap2c}). Its explicit dependence, dictated by the 
renormalization group, reads:
\beqa
 {\cal M}^\prime_1(a)& = & -\frac{(N-1/2)}{(N-1)} \ln a \; , \\ \nonumber
 {\cal M}^\prime_2(a)& = &
\frac{(N-1/2)}{(N-1)^2}\bigl(0.737775-\frac{\pi^2}{96}\bigr)
+\;\frac{(N-1/2)(N-3/4)}{(N-1)^2}\; \ln^2 a \\ \nonumber
& & +\frac{(N-1/2)(N-1/4)}{(N-1)^2}\; \ln a \;.
\hh
\eeqa
It is instructive to examine more closely the differences between
the two forms (\ref{mgap2c}) and (\ref{varM2}): at first sight they look
very different, as not only the functional form of $M_F$,
the power coefficients,
and the perturbative coefficients ${\cal M}_i$, 
${\cal M}^\prime_i$ are different,
but the form of the recursive functions, $f_2$ in equation~(\ref{recur2})
and
$f^\prime_2$ in equation~(\ref{fprime2}) are also different.

Actually, we shall see next that this is in fact
nothing but a RS difference, although in a not too conventional
form. Since, at the second RG order, the RG properties allows
such RS changes, this proves {\it a posteriori} that the construction  
when substitution (\ref{mbarsub}) is made in the relevant 
renormalized
expressions, followed by the appropriate contour integration, gives
an equally acceptable Ansatz. 
In particular it corresponds, up to RS change,
to an expression resumming infinite orders of 
the $x$ series expansion, as is originally aimed.

To make the precise connection between the two different RS choices,
note first that the definition of the variational parameter
$m'''$ in (\ref{mtierce}) and $m''$ in equation~(\ref{mgap2c})
indeed differs (except for $N \to
\infty$):
\beq
m''' = exp\Bigl[{\frac{-1}{2(N-1)}}\Bigr]\; m''\;.
\label{m3m2}
\eeq
The latter relation, by also
comparing~\footnote{From (\ref{m3m2}) one may think at first
that this RS change
is simply given as a specific choice of $a$. This is not so,
since the different form of $f^\prime_2$ in
(\ref{fprime2}) with respect to (\ref{recur2}) 
is also essential, especially for $m''' \to 0$, and
cannot be obtained from the alternative form $f_2$ by simply
changing $a$.} the form of (\ref{fprime2}) with $f_2$  in
(\ref{recur2})
allows to express perturbatively
the relation between the two schemes, for $f^\prime_2$
and $f_2$
sufficiently large:
\beq
\ds{f^\prime_2 \simeq f_2\;\Bigl[1-\frac{1}{2(N-1)}\:\frac{1}{f_2}\;
 +\frac{(3N+1/2)}{4(N-1)^2}
\frac{1}{f^2_2} \;+O\Bigl(\frac{1}{f^3_2}\Bigr) \Bigr] }\;.
\label{relatescheme}
\eeq
One can easily check, by expanding in equation~(\ref{varM2}) in powers of
$1/f_2$ using (\ref{relatescheme}), 
that one
recovers, up to higher order $O(1/f^3_2)$ terms, 
the previous form (\ref{mgap2c})~\footnote{ 
But with $b_2 =\gamma_2 =0$, because in the alternative  
Ansatz
these third order RG coefficients do not appear, while
in the previous Ansatz those are needed at intermediate
stage for consistency, see Appendix B for more details.}.
Consequently the two schemes are {\it perturbatively}
equivalent, as they should.
For $f_2 \to 0$ (equivalently
$f^\prime_2 \to 0$), {\it i.e.} $m'' \to 0$ (equivalently
$m''' \to 0$) the two different
forms have however no
{\it a priori} reason
to give the same results, due to a different behavior of
$f_2$ and $f^\prime_2$ close to the origin. 
 
To proceed, we perform the same type of Pad\'e approximants, as described
in section 3,  with the new form of the mass gap (\ref{varM2}).
Actually  formula
(\ref{varM2}) as it stands is not directly convenient, due to the presence of
an extra branch cut starting at $f^\prime_2 = - b_1/(2b^2_0)$ 
{\it i.e.} on the
positive real axis
($b_1 < 0$), therefore preventing a continuation down to
$m''' \to 0$ like in the calculations of section 3.
One may possibly infer that
$b_1 <0 $ is an artifact of the low orders of perturbation theory ({\it i.e.}
the Gross-Neveu model is asymptotically free to all orders in the large-$N$ 
limit~\cite{GN}). But in any event the analyticity structure
is not uniquely determined, since it clearly depends on
the scheme and precise form of the defining relation for $f_2$, as 
illustrated
here from the two different expressions (\ref{recur2}) and
(\ref{fprime2}). It is in fact possible to recover a
structure
of singularities similar to the previous case 
(while still using
the different information provided from the alternative scheme)
by simply expanding (\ref{varM2}) for
large $f^\prime_2$ (namely in the perturbative range above the extra cut
at $f^\prime_2 = 1/(2(N-1))$,
but keeping the {\it same} definition of the variational
parameter $m'''$ in equation~(\ref{mtierce})\footnote{Of 
course the larger  $N$ is,
the closer one can safely approach the origin for $f^\prime_2$ 
(equivalently $m'''$), 
as expected.}. This replaces expansion~(\ref{varM2}) by
 \beq
\frac{m_F}{\Lambda_{\overline{MS}}}  =
2^{\frac{1}{2(N-1)}} \; \ds \frac{m'''\:a}{2i\pi} \;
\oint\frac{du{\rm e}^{u}}{f_{2}^{\frac{N-1/2}{N-1}}(u)}
\Bigl\{ 1+\frac{{\cal M''}_{1}(a)}{f_2}+\frac{{\cal
M''}_{2}(a)}{f_{2}^{2}}+\ldots \Bigr\} \hh ,
\label{mgap2d}
\eeq
where $f_2$ verifies the very same relation as in equation~(\ref{recur2}) 
(but
now with
$m'' \to m'''$), and the cut again starts at $f_2 = 0$ towards the real
negative axis. The perturbative coefficients are of course different:
\beqa
{\cal M''}_{1}(1) & = & \frac{N-1/2}{4 (N-1)^2} \hh , \nonumber \\
{\cal M''}_{2}(1) & = & {\cal M}'_2(1)-(N-1/2)
\frac{(16N^2-14N-3)}{64(N-1)^4}
\;.
\eeqa
One can now directly apply the Pad\'e techniques as described in section
3, to the form (\ref{mgap2d}). The most interesting 
results are illustrated with the Pad\'e~(2,3)
case in the last line of table 2, and are substantially better than
the corresponding order results in the previous scheme. In particular
the error is much more stable when varying $N$. Optimal values 
in this alternative scheme for
Pad\'e approximants of lower orders 
are only slightly different from those of the original
scheme as given in table 2.
For larger $N$, the two schemes
give more and more similar results, as expected since for $N \to \infty$
they are {\it strictly} equivalent: $m''' \to m''$, and
all perturbative coefficients ${\cal M}_i(1)$ go  to 0.
These  different facts provide a good
check of the stability of our results with respect to changes in the
renormalization scheme. Indeed it strongly indicates that it should 
also be possible to optimize with respect to the RS scheme (in addition
to the optimization with respect to $a$).
The variation of
the scheme through the parameter $a$ 
(whose dependence does differ in the alternative Ansatz 
at the next-to-leading order,  $(\ln a)/f^2_2$), provides an extra
 illustration of the consistency of our
results. Besides, these two different renormalization 
schemes are important to
clarify the variational expansion procedure when applied either to bare
or renormalized parameters.
To illustrate a more general RS dependence one could further exploit 
the RS dependence of the \RG\
coefficients themselves, {\it e.g.} starting with $\gamma_1$.
Although we have not tried explicitely this in the present case, we see
no compelling reasons to study a more general case within the framework
of the GN model, where we know the exact results anyway.
The preceding facts 
should already be considered a strong indication that the variational
mass method does lead to non-trivial results, at least in 
appropriate renormalization schemes.

The vacuum energy density can be treated along the
same lines. However, in that case, the change of sign of the energy
when $m''$ approaches $0$ complicates matters in a way which we do not report
here, as it is a separate issue.

\section{Conclusion}
 
In this paper, we have shown how to introduce a variational
procedure in an asymptotically free field theory, in a way which
is compatible with the renormalization group: we have found
how to rescale the variational parameter in order to take account of
all infinities (at least in the framework of dimensional regularization)
so that the physics of the variational approximation is smooth when
the cut-off is removed (in our case, when the dimension is varied
around its critical value). We have further shown that the  numerical
results obtained with an extra very simple extrapolation procedure
 are in good to remarkable agreement with exact values in the case
of the GN model.
This establishes the theoretical possibility and  potential  usefulness
of perturbative calculations for computing non-perturbative quantities.
 This framework is in principle directly applicable
to the QCD case, apart form possible complications due to the
{\em a priori} different analyticity structure of $m$ 
of the relevant QCD expressions. 
In a subsequent paper~\cite{AGKN},
using a bare chiral symmetry breaking fermion mass as
variational parameter, we shall  obtain values for some
parameters of spontaneous chiral symmetry breaking,
$m_Q$ (the part of the constituent quark mass due to chiral
symmetry breaking), the pion decay constant $f_\pi$ and
the condensate $<\bar{\psi} \psi >$. \\
\newpage
{\large \bf Acknowledgments} \\
 
One of us (A. N.) is grateful for their hospitality to the
Lawrence Berkeley Laboratory,
where this work was supported in part by the Director,
Office of Energy research, Office of High Energy and
Nuclear Physics, Division of High Energy Physics of
the U.S. Department of Energy under Contract
DE-AC03-76SF00098 and to the Physics Department,
University of California, Berkeley, where this work was completed
with partial support from National Science Foundation
grant PHY-90-21139. C. A. is grateful to the theory group of Imperial
College for their  hospitality and acknowledges financial
support from the Minist\`{e}re de la Recherche et de la Technologie,
and from E.E.C. grant No. ERBCHBICT941235. J.-L. K. acknowledges
support from a CERN fellowship.

\appendix
\section{A contour integral resumming the $x$ dependence}
\setcounter{equation}{0}
We shall explain here in some details how to resum the $x$ series
generated from the perturbative expansion of (\ref{LGN}).
Consider the one-loop RG invariant (bare) expression for the
mass $m_F$, as given in eq. (\ref{mf1}):
\beq
m_F  = \frac{\mo}{(1- 4\pi b_0 \go^2
\ggt \mu^{-\eps} m^{\eps}_F(\mo) )^{\frac{\gamma_0}{2b_0} } }
\label{Mbareansatz11}
\eeq
where the substitution
\beq m_0 \to m_0 (1-x); ~~~~g^2_0 \to g^2_0 x,
\label{substitut2}
\eeq
provides a new quantity $m_F(x)$.
To pick up the $x^q$ order term in $m_F(x) \equiv \sum^\infty _{q=0}
a_q x^q $ (having in mind
that we are actually interested in the limit $x \to 1$),
a convenient trick is by contour integration:
\beq
m^{(q)}_F (x \to 1) \to \sum^q _{k=0} a_k =
{1\over{2 i \pi}} \oint dx (\frac{1}{x}+\cdots
+\frac{1}{x^{q+1}})\;m_F(x)
.
\label{contour1}
\eeq
Now performing the sum in (\ref{contour1}) exhibits a $(1-x)^{-1}$ factor
which cancels the $(1-x)$ from (\ref{substitut2}).
This results in the expression~\footnote{In (\ref{contour2})
there appeared in fact a factor of $1- x^{-(q+1)}$, from which only the last
term contributes to the integral due to the analyticity of
$f_0(x)$ defined in (\ref{f0def}).}:
\beq
m^{(q)}_F (x \to 1) \to
{1\over{2 i \pi}} \oint dx x^{-(q+1)}\;m_0
[f_0(x)]^{-\frac{\gamma_0}{2b_0}}\;,
\label{contour2}
\eeq
where the contour is counterclockwise around the origin, and
for convenience we defined the (recursive) function
\beq
f_0(x) \equiv 1- 4\pi b_0 \;x\; \go^2 \ggt \; 
m^{\epsilon}_0 (1-x)^{\epsilon}\; (f_0)^{-\epsilon \frac{\gamma_0}{2b_0}}\;,
\label{f0def}
\eeq
directly dictated from eq. (\ref{Mbareansatz11}).
$f_0(x)$ has (evidently) a power series expansion in $x$, but
admits also an expansion in $(1-x)$, as noted by
inverting its defining relation (\ref{f0def}). This implies,
 in particular,
that
$x =1$ is an (isolated) pole of $m_F$. \\
{\em Provided} that no extra singularities lie in the way, one may distort
the integration contour in (\ref{contour2})
to go around the cut lying along the
real positive axis and starting at $x =1$.
Actually one can go a step further and reach the $q \to \infty$ limit:
after distorsion of the contour only the vicinity of $x =1$ survives
for $q \to \infty$, that one can analyse by changing variable to
\beq
1 -x \equiv {v \over q}
\eeq
and rescaling $m_0$ by introducing $m_0 = m^{'}_0 q$
(keeping $m^{'}_0$ fixed as
$q$ goes to infinity). One finds in place of (\ref{contour2})
\beq
m^{(q)}_F(q \to \infty) \to {1\over{2 i \pi}} \oint {dv \over v} e^v
\; { v\; m^{'}_0 \over{f_0(v)^{\gamma_0
\over{2b_0}}} }
\label{contour3}
\eeq
where now $f_0(v) \equiv 1- 4\pi b_0 \; g0^2 \ggt
(m_0 v)^{\epsilon} (f_0)^{-\epsilon \frac{\gamma_0}{2b_0}}$.
The crucial point in (\ref{contour3}) is, that
once performing renormalization
via $m_0 = \bar m Z_m$, $g^2_0 = \bar \mu^{-\epsilon} Z_g
g^2$,   $m_F$
is finite to all orders:
\beq
m_F = {1\over{2 i \pi}} \oint dv e^{v} {\bar m \over {f^{{\gamma_0
\over{2b_0}}}}}
\label{contour5}
\eeq
where the
renormalized function $f = Z_g f_0 = 1 +2b_0 \bar g^2 \ln [(\bar m v/\bar \mu)
\;f^{-(\gamma_0/2b_0)}\;]$.
 
We have thus recovered finite quantities with a non-trivial
$x$ expansion. In the latter derivation 
we only included the one-loop RG dependence, which 
is the exact result in the large $N$ limit only.
For arbitrary $N$ one should include
in the derivation the non-logarithmic
perturbative terms,  present e.g at the two loop order 
in (\ref{mpertfinite}).
This can be done 
without affecting the contour integration
properties, except that the resulting expression of $m_F$ has a more
complicated structure around $v \simeq 0$, but can be systematically
expanded around the origin in the way discussed in section 2.
Generalization of the previous construction to
the next RG order is straightforward, since the recursive
function $f$ has a very similar form as in (\ref{f0def}) above, where
only the power coefficients are changed.

\section{Bare RG Ansatz}
\setcounter{equation}{0}
 
We give here some useful expressions 
needed for the construction of the generalized Ansatz using third order
RG functions, as introduced in section 2.3, and leading to our main
results
(\ref{mresult}), (\ref{mcoef}) and (\ref{mgap2c})
for the mass gap. Derivation of
the vacuum energy (\ref{energ2c}) is very similar, apart from
the slight complication due to the subtracted terms as discussed at
the end of section 2. 

We start from expression (\ref{mf2}), where
the constants $A$, $B$ and $D$ are first to be determined by requiring
equation~(\ref{mf2}) to be finite as $\epsilon \to 0$, when expressed
in terms of $m$ and $g^2$ with the help of RG mass and coupling
counterterms at a given order (and in a specific renormalization scheme).
The coefficient $\lambda$, $\rho, \ldots$ are then obtained by matching
the perturbative expansion of~(\ref{mf2}) to the perturbative (bare)
expression for $m_F$, eq.~(\ref{ptg4}). 

Defining the \RG\ coefficients governing the coupling constant and mass
evolution, respectively, as
\beq
\beta(g) \equiv \mu {dg\over{d \mu}} = \frac{\eps}{2} g
-b_0~ g^3 -b_1~ g^5 -b_2~ g^7 -\ldots
\label{beta}
\eeq
and
\beq
\gamma_m(g) \equiv -{\mu \over m}{dm \over{d \mu}} =
\gamma_0~ g^2 +\gamma_1~ g^4 +\gamma_2~ g^6 +\ldots
\label{gamma}
\eeq
the coefficients $b_i$ and $\gamma_i$, known in the $\overline{MS}$
scheme
up to three loop order~\cite{GRAC} in the Gross-Neveu model, are expressed as
\beq
\label{bi}
b_0 = \frac{N-1}{2\pi},\;b_1 = -\frac{N-1}{4\pi^2}, \;
b_2 = -\frac{(N-1)(2N-7)}{32\pi^3} ;
\eeq
and
\beq
\label{gi}
\gamma_0 = \frac{N-1/2}{\pi},\; \gamma_1 = -\frac{N-1/2}{4\pi^2}, \;
\gamma_2 = -\frac{(N-1/2)(4N-3)}{16\pi^3} .
\eeq
 From these one obtains explicit expressions of the counterterms, 
$Z_g$ and $Z_m$, as expansions both in $g^2$ and $\eps$ to the required
order. 
We need a counterterm for ${\cal F}$ in (\ref{mf2}) 
as well, that we define according
to $ 
{\cal F}_0 \equiv Z_F {\cal F}
$,
where $Z_F$ can be determined consistently together with $A$, $B$ and $D$
for given $Z_m$ and $Z_g$.
In terms of $Z_g$ and $Z_m$, the finiteness of eq.({\ref{mf2}) requirement 
explicitely reads
\beqa
\label{finiteness}
& B \;\ln z +\ln Z_m \equiv \; {\rm finite}; \\ \nonumber
& (1\:+\eps A)\; \ln z +\eps\: \ln Z_m + \ln Z_g \equiv \; {\rm finite};
\\ \nonumber
& D \equiv  1+\eps A \; ;
\eeqa
where we exhibited the singular part of $Z_F$ as
\beq
\ln z \equiv \ln \Bigl[1 +\frac{g^2}{2\pi}
-\frac{(N-1)g^2}{\pi \eps}\Bigr].
\label{singular}
\eeq
 From inspection of the formal expansion of
 (\ref{mf2}) it turns out that actually $A$ is needed to $O(\eps)$ 
and $B$ to $ O(\eps^2)$, to fix unambiguously
the finite coefficients  $\lambda$ and $\rho$ in (\ref{F}) by
matching with~(\ref{ptg4}). Accordingly from
(\ref{finiteness}) one needs to know $Z_m$, $Z_g$ to order
$O(1/\eps^3)$: i.e  
the third order RG coefficients $b_2$,
$\gamma_2$ in (\ref{bi}), (\ref{gi}) should be 
included for consistency.

The required 
expansion in powers of $1/\eps$ of $\ln Z_g$, $\ln Z_m$ and $\ln z$
is completely determined by
the expansion in $\eps$ of the perturbative zero, $g_P$, of the
beta function, which is easily derived as
\beq
g^2_P = \ds{\frac{\eps}{2b_0}-\frac{b_1}{4b^3_0}\eps^2 +
(\frac{b^2_1}{4b^5_0}-\frac{b_2}{8b^4_0})\eps^3 +O(\eps^4)}\;.
\eeq
In this way we obtain after straightforward algebra,
\beqa
\label{AandB}
 A & = &\ds{\frac{N}{N-1} +\frac{N-1/2}{(N-1)^2}\;\eps +O(\eps^2)}\; ,
\\ \nonumber
 B & = & \ds{\frac{N-1/2}{N-1} +\frac{3}{4}\frac{N-1/2}{(N-1)^2}\;\eps
+\frac{N(N-1/2)}{8(N-1)^3}\;\eps^2 +O(\eps^3) }\; ,
\eeqa
and~\footnote{Note that the
order $\eps$ term in $\lambda$ is required to determine $\rho$.}
\beqa
\label{lambdamu}
  \lambda & = & \frac{3}{4\pi}\frac{N-1/2}{N-1} \\ \nonumber
& \;  & +\frac{N-1/2}{8\pi(N-1)}
\Bigl[\frac{N}{N-1}+3(\gamma_E -\ln 4\pi)\Bigr]\; \eps \;+O(\eps^2)\; ,
\\ \nonumber
 \rho & = &
 - \frac{(N-1/2)(36N^2-62N+17)}{64\pi^2(N-1)^2} \\ \nonumber
 & \; &  -\frac{N-1/2}{2\pi^2} \bigl(-0.725551\ldots + \frac{\pi^2}{48}\bigr) 
+O(\eps) \; , 
\hh
\eeqa
which directly leads to expressions (\ref{mcoef}) for $\eps \to 0$. 
Let us finally remark that (\ref{finiteness}) 
leaves over non-zero finite perturbative terms, which are
however 
completely determined order by order. This is inherent to our scheme
here where the counterterm for ${\cal F}$
could not be put into a form where only the singularities
appear (like is the case, by definition, for $\ln Z_m$ and $\ln Z_g$
in the $\overline{MS}$ scheme). Explicitely 
after using (\ref{AandB}) we obtain
\beq
\label{remnant}
B\; \ln z +\ln Z_m = -\frac{(N-1/2)}{(N-1)}\frac{g^2}{4\pi} +
\frac{(N-1/2)(N+1)}{(N-1)}\frac{g^4}{16\pi^2}+O(g^6)\; ,
\eeq
which obviously gives perturbative corrections
when re-expanding expression~(\ref{mresult}),
and which are indeed necessary to recover consistently the
perturbative expression for the 
renormalized mass~(\ref{mpertfinite})~\footnote{The third 
order renormalization
group dependence, via $b_2$ and $\gamma_2$ in~(\ref{AandB}), (\ref{lambdamu})
and (\ref{remnant}) actually cancels 
in the (two-loop) perturbative renormalized
mass expression~(\ref{mpertfinite}).}.

\newpage
\centerline{\large \bf Figure Captions.}
 
\vspace{20pt}
 
\noindent
Figure 1. Mass gap graphs at order 2.
 
\noindent
Figure 2. Vacuum energy graphs at order 2.
 
\noindent
Figure 3. Mass gap ratio $m_{F}/\Lambda_{\overline{MS}}$ versus variational
parameter $m''$, for values of the scaling parameter $a$ ranging
from $a=1.1$ (lowest curve) to $a=1.2$ (highest curve).
The horizontal dashed curve represents the exact value.
We have taken $N=3$ for the number of flavors.
 
\noindent
Figure 4. Vacuum energy $E_{0}/{\Lambda_{\overline{MS}}}^2$ versus $m''$,
for values of $a$ ranging from $a=0.9$ (lowest curve)
 to $a=1.1$ (highest curve). The horizontal dashed curve
gives the exact value. The number of flavors is $N=3$ .
 
\noindent
Figure 5. 
Comparison of different Pad\'e approximant types and orders,
versus exact
$m_{F}/\Lambda_{\overline{MS}}$, for
$N=2$. The curves showing extrema at 2.27, 1.62, 1.76, 1.93 and 1.87,
have been obtained from the $p_{1,1}(u)$,
$p_{2,3}(u)$, $p_{2,2}(u) \times e^{(c1/u)}$,
$p_{1,2}(u) $ and $p^{'}_{2,3}(u)$
Pad\'e approximants, respectively. 
\newpage
 
\begin{figure}[t]
\epsfxsize=9truecm
\epsfysize=5truecm
\centerline{\epsffile{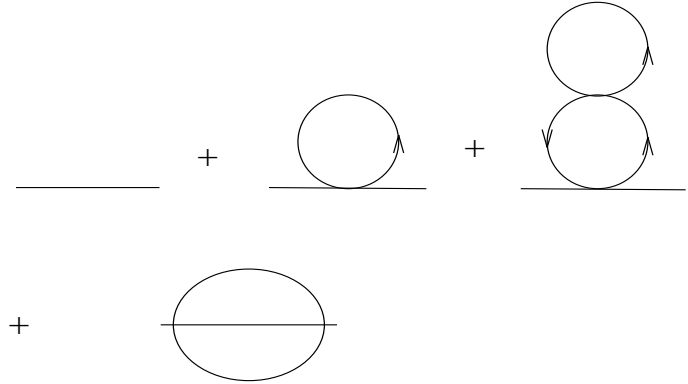}}
\vspace{1.cm}
\caption{Mass gap graphs at order 2.}
\label{fig1}
\end{figure}
 
\vspace{60pt}
 
\begin{figure}[t]
\epsfxsize=6truecm
\epsfysize=6truecm
\centerline{\epsffile{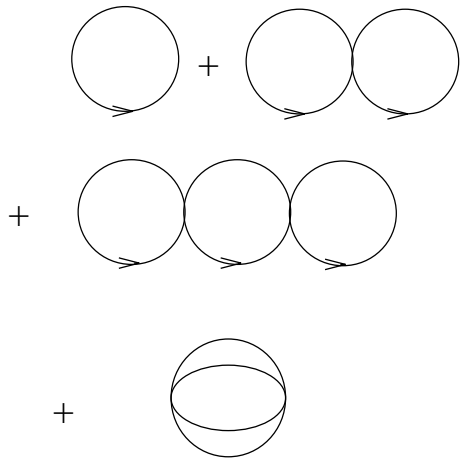}}
\vspace{1.cm}
\caption{Vacuum energy graphs at order 2.}
\label{fig2}
\end{figure}
 
\begin{figure}[t]
\epsfxsize=15truecm
\epsfysize=15truecm
\centerline{\epsffile{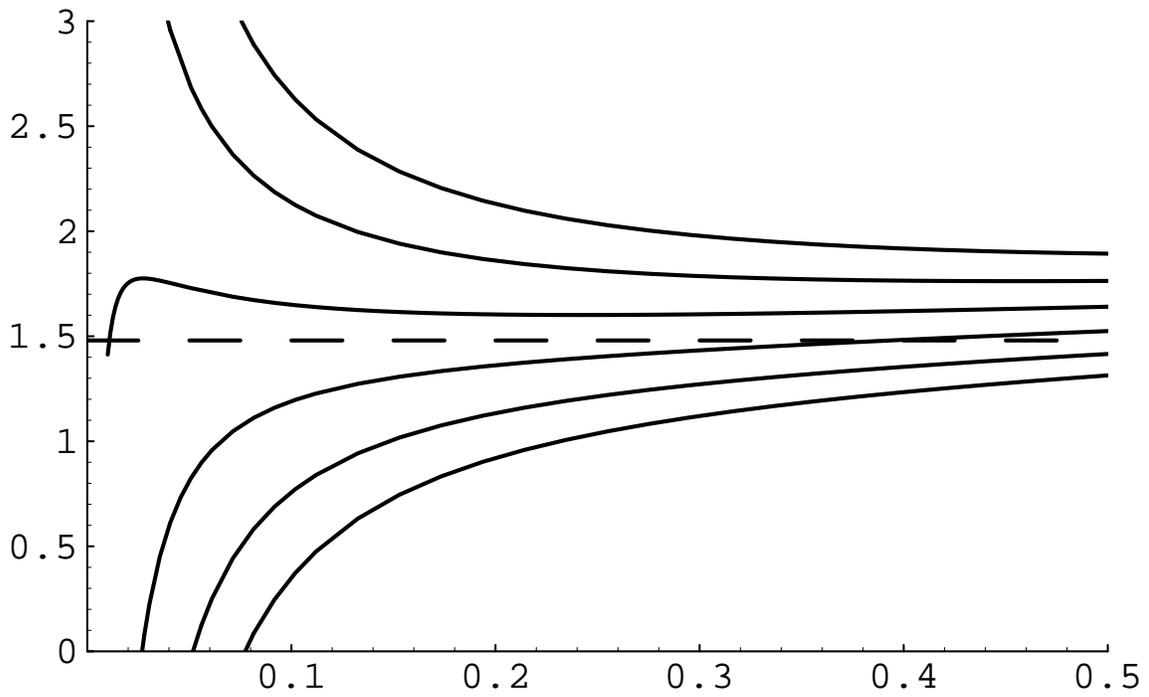}}
\vspace{-1.cm}
\caption{Mass gap $m_{F}/\Lambda_{\overline{MS}}$ 
versus variational parameter
$m''$ .}
\label{fig3}
\end{figure}
 
\begin{figure}[t]
\epsfxsize=15truecm
\epsfysize=15truecm
\centerline{\epsffile{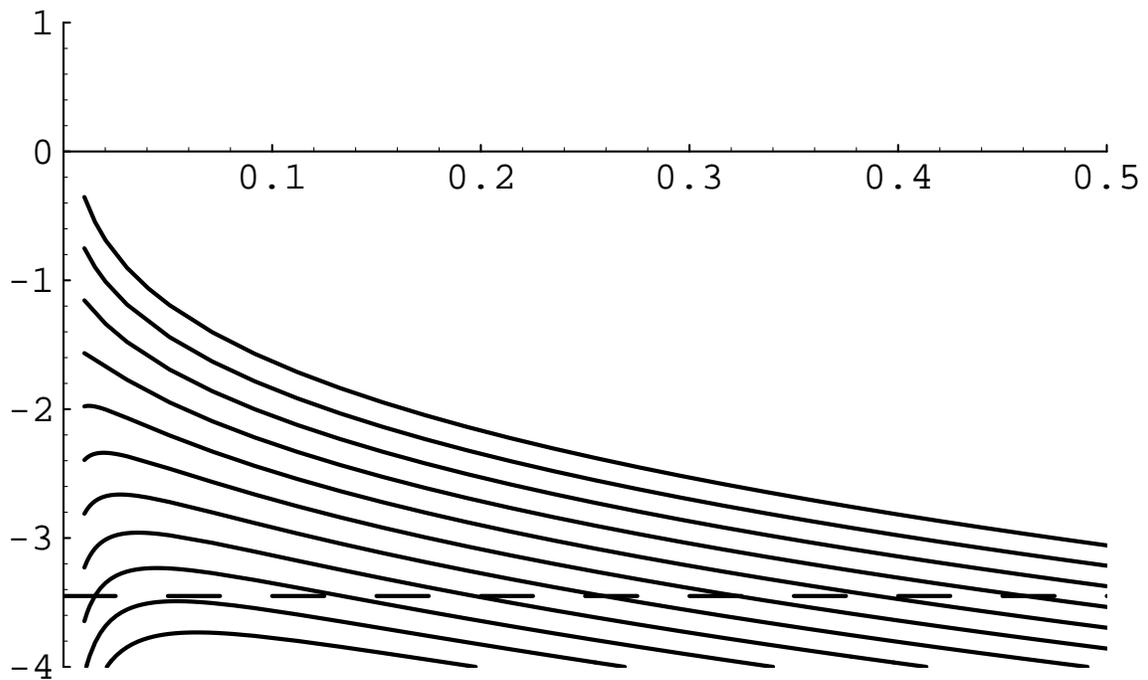}}
\vspace{-1.cm}
\caption{Vacuum energy $E_{0}/{\Lambda_{\overline{MS}}}^2$ versus variational
parameter $m''$.}
\label{fig4}
\end{figure}

\begin{figure}[t]
\epsfxsize=15truecm
\epsfysize=15truecm
\centerline{\epsffile{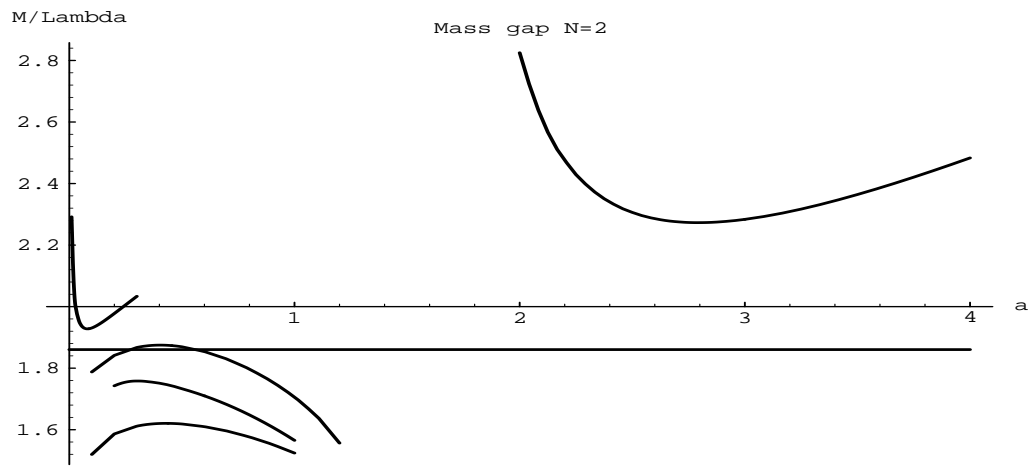}}
\vspace{-1.cm}
\caption{Comparison of different Pad\'e approximant types and orders,
versus exact 
$m_{F}/\Lambda_{\overline{MS}}$, for
$N=2$.} 
\label{fig5}
\end{figure}
 
\end{document}